%% file: main.tex
\documentclass[sigconf]{acmart}

\AtBeginDocument{%
  \providecommand\BibTeX{{%
    \normalfont B\kern-0.5em{\scshape i\kern-0.25em b}\kern-0.8em\TeX}}}

\setcopyright{acmcopyright}
\acmConference[SIGIR '21]{Proceedings of the 44th International ACM SIGIR Conference on Research and Development in Information Retrieval}{July 11--15, 2021}{Virtual Event, Canada}
\acmBooktitle{Proceedings of the 44th International ACM SIGIR Conference on Research and Development in Information Retrieval (SIGIR '21), July 11--15, 2021, Virtual Event, Canada}
\acmPrice{15.00}
\acmDOI{10.1145/3404835.3462789}
\acmISBN{978-1-4503-8037-9/21/07}

\usepackage{graphicx}
\usepackage{booktabs}
\usepackage{caption}
\usepackage{multirow}
\usepackage{subcaption}
\usepackage{bbding}
\usepackage{array}
\usepackage{balance}

\begin{document}

\title{OpenMatch: An Open Source Library for Neu-IR Research}
\author{Zhenghao Liu$^{\heartsuit*}$, Kaitao Zhang$^{\heartsuit*}$, Chenyan Xiong$^{^\spadesuit}$, Zhiyuan Liu$^\heartsuit$ and Maosong Sun$^\heartsuit$}\thanks{$*$\ \ indicates equal contribution.}
\affiliation{Tsinghua University$^\heartsuit$, Microsoft Research$^\spadesuit$} 
\affiliation{
\texttt{\{liu-zh16, zkt18\}@mails.tsinghua.edu.cn};
\texttt{chenyan.xiong@microsoft.com};
\texttt{\{liuzy, sms\}@tsinghua.edu.cn}
}

\input{Sections/0_Abstract}


\maketitle

\input{Sections/1_Introduction}
\input{Sections/3_System_Overview}
\input{Sections/4_OpenMatch_Library}
\input{Sections/5_Model_Hub}
\input{Sections/2_Related_Work}

\input{Sections/6_Conclusion}
\input{Sections/7_Acknowledge}
\balance
\bibliographystyle{ACM-Reference-Format}
\bibliography{citation}

\end{document}

%% file: Sections/0_Abstract.tex
\begin{abstract}
OpenMatch is a Python-based library that serves for Neural Information Retrieval (Neu-IR) research. It provides self-contained neural and traditional IR modules, making it easy to build customized and higher-capacity IR systems. In order to develop the advantages of Neu-IR models for users, OpenMatch provides implementations of recent neural IR models, complicated experiment instructions, and advanced few-shot training methods. OpenMatch reproduces corresponding ranking results of previous work on widely-used IR benchmarks, liberating users from surplus labor in baseline reimplementation. Our OpenMatch-based solutions conduct top-ranked empirical results on various ranking tasks, such as ad hoc retrieval and conversational retrieval, illustrating the convenience of OpenMatch to facilitate building an effective IR system. The library, experimental methodologies and results of OpenMatch are all publicly available at \url{https://github.com/thunlp/OpenMatch}.

\end{abstract}

%% file: Sections/1_Introduction.tex
\section{Introduction}
With the rapid development of deep neural networks, Information Retrieval (IR) shows better performance and benefits lots of applications, such as open-domain question answering~\cite{chen2017reading} and fact verification~\cite{thorne2018fact}.
Being neural has become a new tendency for the IR community, which helps to overcome the vocabulary mismatch problem that comes from sparse retrieval models. Neu-IR models show their effectiveness in implementing both retrieval and reranking modules to build IR systems.
\input{Tables/package}

Recent research comes up with lots of IR models and has to compare with lots of baseline models, such as Neu-IR models and sparse retrieval methods. However, lots of baseline models have no standard implementations and experimental details, which forces some work to borrow empirical results from related work and consumes lots of efforts from researchers to reproduce the results of these methods, even for some feature-based learning-to-rank methods~\cite{dai2018convolutional,dai2019deeper}, becoming an obstacle for some IR researchers.

This paper proposes OpenMatch, an open-source library designed to support Neu-IR research. As shown in Table~\ref{tab:toolkit}, with well-defined user interfaces, the OpenMatch library provides various modules to implement different ranking models, such as sparse retrievers~\cite{lavrenko2017lm,robertson2009probabilistic}, dense retrievers~\cite{karpukhin2020dense,xiong2020approximate} and neural rerankers~\cite{xiong2017end,dai2018convolutional,qiao2019understanding}, making it easy to build a tailored retrieval pipeline by choosing neural/non-neural ranking models and combining ranking features from different models. OpenMatch is built on Python and Pytorch to maintain well system encapsulation and model extensibility. Our OpenMatch library incorporates some few-shot training modules, which implement some advanced Neu-IR training strategies targeted at dealing with the data scarcity problem in lots of ranking scenarios~\cite{hawking2004challenges,chirita2005using,arora2018challenges,roberts2020trec}. The few-shot training strategies consist of two kinds of methods: leveraging domain knowledge in ranking~\cite{liu2018entity}; synthesizing and reweighting large-scale weak supervision signals to train Neu-IR models~\cite{zhang2020selective,sun2020meta}.

OpenMatch also provides experimental results of various baselines on ranking benchmarks, which are widely used in different ranking scenarios, such as ad hoc retrieval~\cite{qiao2019understanding,dai2019deeper}, evidence retrieval in question answering~\cite{chen2017reading,thorne2018fact} and conversational search~\cite{qu2020open,cast2020overview}. OpenMatch implements several widely-used Neu-IR models and reproduces the ranking results of corresponding work. We submit the reimplemented results to document and passage ranking tasks on the MS MARCO leaderboard\footnote{Team name: THU-MSR~\url{https://microsoft.github.io/msmarco/}} to further guarantee the reproducibility with the hidden test. Besides, our OpenMatch based solutions rank first\footnote{Team name: CMT~\url{https://ir.nist.gov/covidSubmit/archive.html}} of the automatic group in TREC COVID Round 2, demonstrating its ability to build an effective IR pipeline system for a new ranking problem that has few training data.


%% file: Tables/package.tex
\begin{table}[tbp]
    \centering
    \caption{The Comparison of Different IR Toolkits.}\label{tab:toolkit}
    \begin{tabular}{l|c|c|c|c}
    \hline
    \multirow{2}{*}{Toolkit} & \multicolumn{2}{c|}{Retriever} & Neural &  Few-Shot\\
    \cline{2-3}
    & Sparse & Dense & Reranker & Training \\
    \hline
    Anserini~\cite{yang2017anserini} & \CheckmarkBold & & & \\
    MatchZoo~\cite{guo2019matchzoo} & & & \CheckmarkBold & \\
    OpenNIR~\cite{macavaney2020opennir} & \CheckmarkBold & & \CheckmarkBold & \\
    Capreolus$^{\ref{url:cap}}$ & \CheckmarkBold & & \CheckmarkBold & \\
    Pyserini~\cite{lin2021pyserini} & \CheckmarkBold & \CheckmarkBold &  & \\
    \hline
    OpenMatch & \CheckmarkBold & \CheckmarkBold & \CheckmarkBold & \CheckmarkBold \\
    \hline
    \end{tabular}
\end{table}

%% file: Sections/3_System_Overview.tex
\section{System Overview}

\begin{figure}[tbp]
    \centering
	\includegraphics[width=0.925\linewidth]{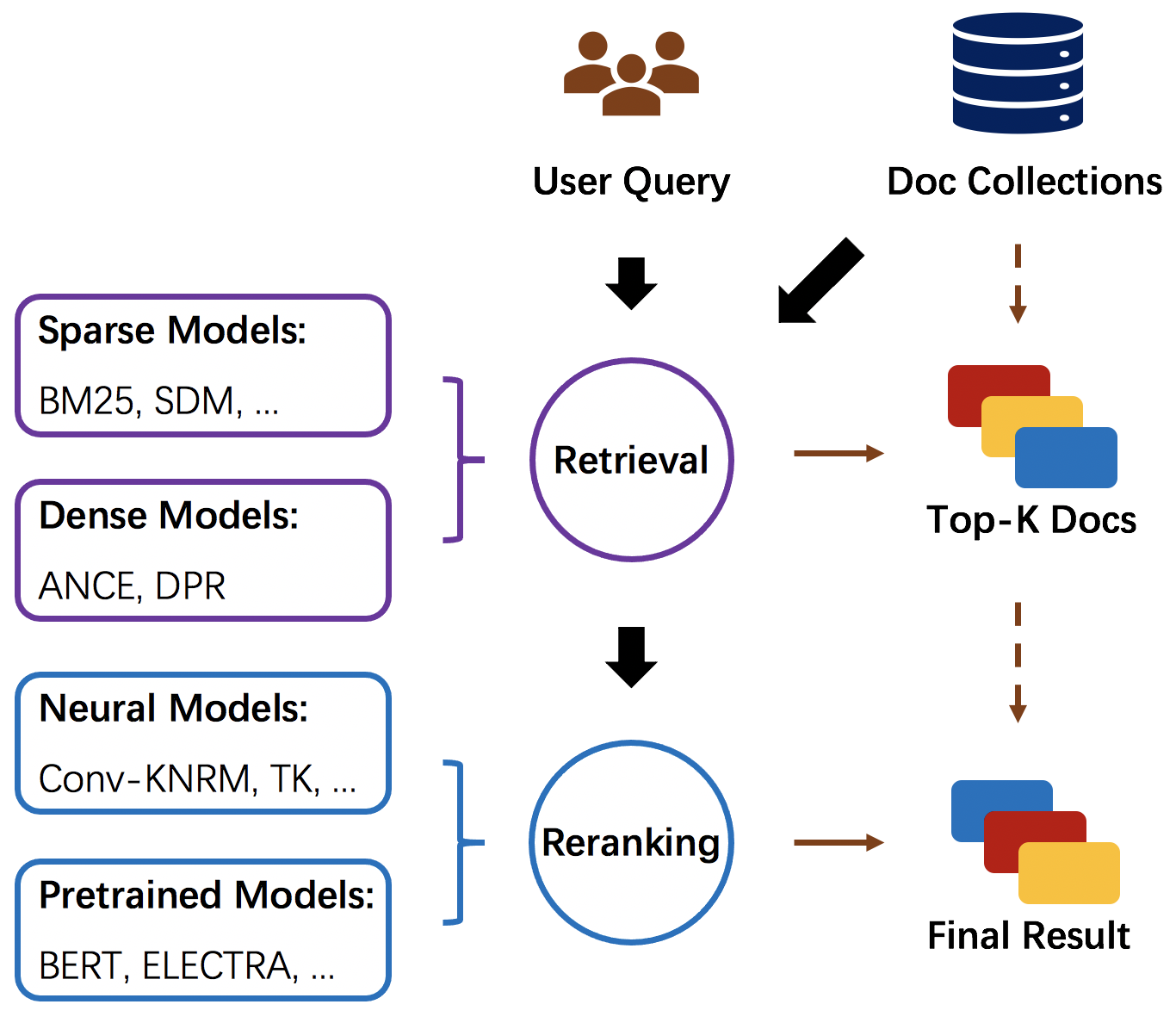}
	\caption{The Two Step Retrieval Pipeline of OpenMatch.}
	\label{fig:pipeline}
\end{figure}

\begin{figure}[tbp]
    \centering
    \begin{subfigure}{0.267\textwidth}
         \centering
         \includegraphics[width=\textwidth]{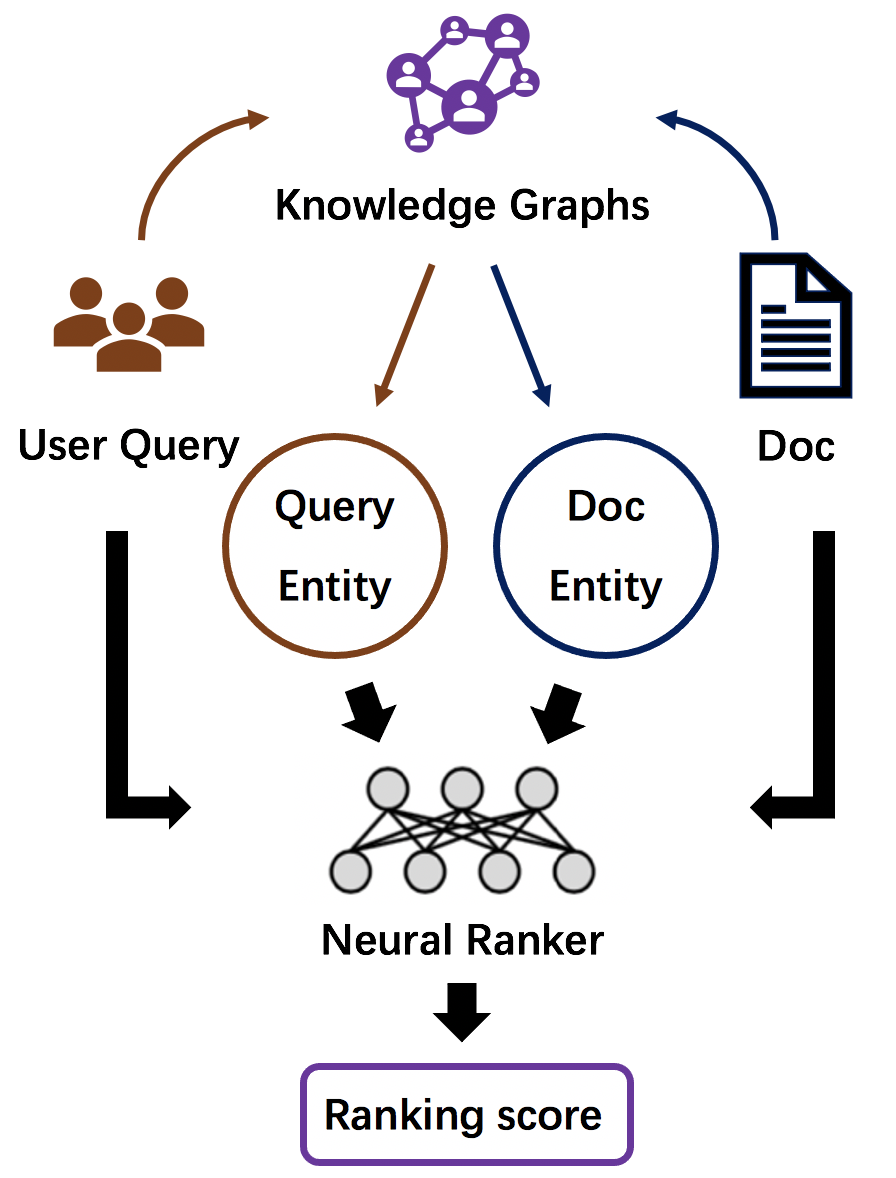}
         \caption{Knowledge Enhancement.}
    \end{subfigure}
    \hfill
    \begin{subfigure}{0.205\textwidth}
         \centering
         \includegraphics[width=\textwidth]{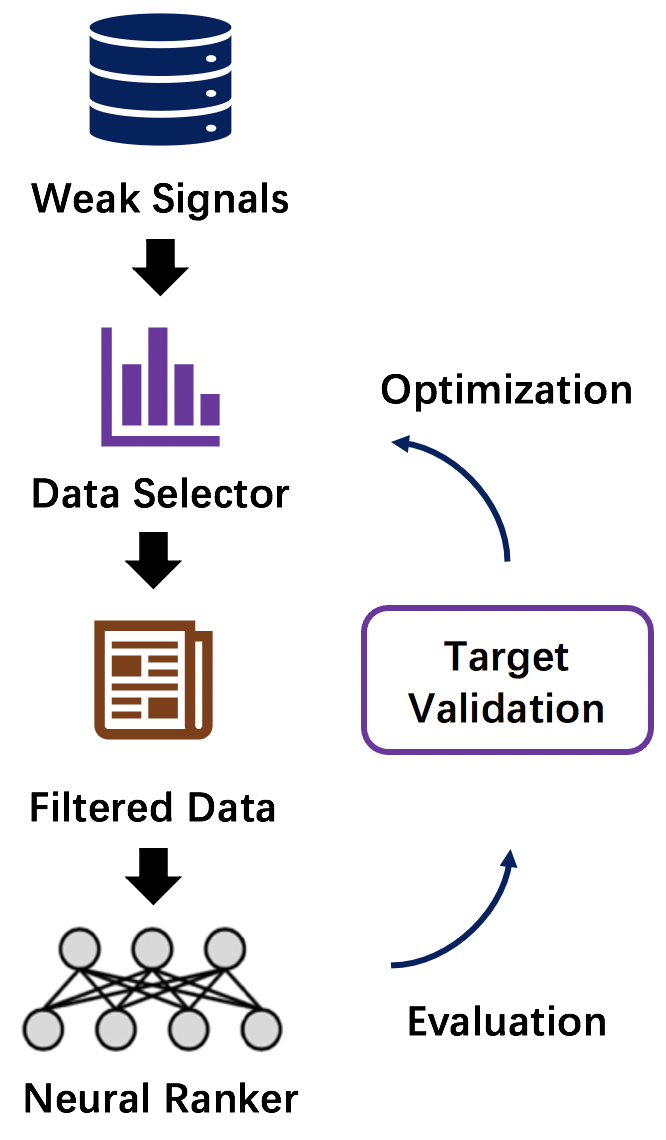}
         \caption{Data Augmentation.}
    \end{subfigure}
\caption{The Few-Shot Training Methods of OpenMatch.}
\label{fig:adaption}    
\end{figure}

OpenMatch provides several pre-defined modules and helps users build an effective IR system with retrieval and reranking steps, as shown in Figure~\ref{fig:pipeline}. In the retrieval stage, we first inherit Anserini~\cite{yang2017anserini} and ANCE~\cite{xiong2020approximate} to build the document index for sparse retrieval and dense retrieval to efficiently search candidate documents from a large-scale document collection. Then, for the reranking stage, OpenMatch provides various modules to extract ranking features with sparse retrieval models, dense retrieval models, Neu-IR models, and pretrained IR models. The neural modules in OpenMatch are defined to conveniently implement different Neu-IR models. Moreover, as shown in Figure~\ref{fig:adaption}, OpenMatch incorporates several advanced training methods to further broaden the advance of Neu-IR models in few-shot ranking scenarios. These few-shot training methods mainly focus on alleviating data scarcity in ranking scenarios by leveraging domain knowledge and data augmentation. Besides, our library also provides some modules for data preprocessing, ranking feature ensemble and automatic evaluation.

%% file: Sections/4_OpenMatch_Library.tex
\section{OpenMatch Library}
\label{sec:library}

\begin{figure*}[tbp]
    \centering
    \begin{subfigure}{0.499\textwidth}
         \centering
         \includegraphics[width=\textwidth]{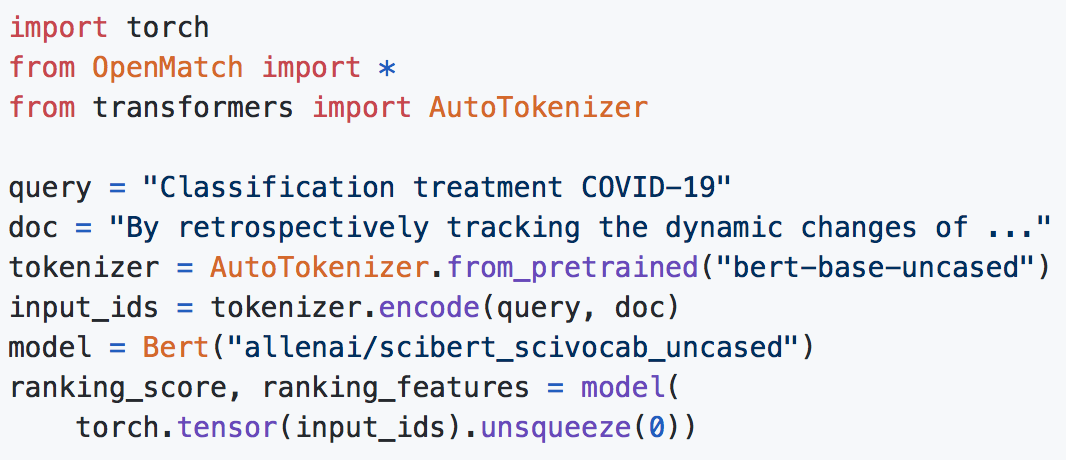}
         \caption{Pretrained Models.}
    \end{subfigure}
    \hfill
    \begin{subfigure}{0.497\textwidth}
         \centering
         \includegraphics[width=\textwidth]{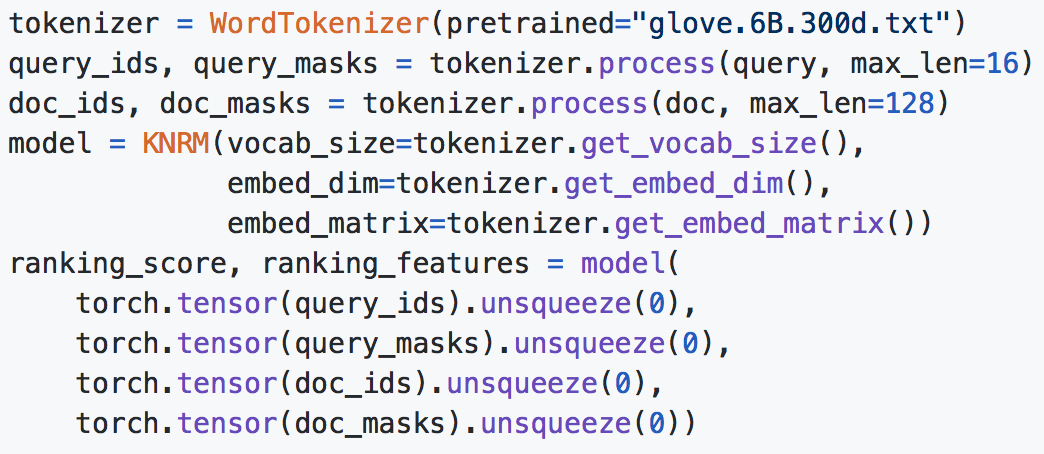}
         \caption{Neural Models.}
    \end{subfigure}
\caption{OpenMatch Quick Start Examples.}
\label{fig:example}    
\end{figure*}

OpenMatch is built on PyTorch and provides several extensible modules in the whole IR pipeline for data processing, Neu-IR model implementation, few-shot training, model inference, and ranking performance evaluation.

\subsection{Data Processing}
This part presents the data processing modules in OpenMatch that can be used to deal with various data formats for ranking and develop customized classes for individual data processing, including \texttt{tokenizer}, \texttt{dataset}, and \texttt{dataloader}.

\textbf{Tokenizer.} OpenMatch follows HuggingFace's Transformers~\cite{Wolf2019HuggingFacesTS} to design the \texttt{Tokenizer} class for Neu-IR models, which makes it easy-to-use and extensible. The \texttt{Tokenizer} can be initialized from raw vocabularies and word embeddings. It maintains a token-to-id map to convert raw texts into token ids for Neu-IR models and provides two functions, stopword removal and word stemming, to process words, which are widely used in IR models. For pretrained language models, our \texttt{Tokenizer} module inherits the same tokenizer implementation from HuggingFace's Transformers library.

\textbf{Dataset.} The \texttt{Dataset} class in OpenMatch is inherited from the \texttt{Dataset} class of PyTorch. It provides several functions to process the raw data with various formats for pair-wise training, point-wise training, and model inference. The collation function of OpenMatch further supplies an API to stack data instances into mini-batches for the \texttt{Dataloader} module.

\textbf{Dataloader.} OpenMatch also inherits the \texttt{Dataloader} class of PyTorch to generate batched data for Neu-IR models. It is convenient to set batch size, shuffle flags, and the number of workers in loading data. With the \texttt{Dataloader} module of PyTorch, OpenMatch can be easily extended to model inference process, single-GPU training process and multi-GPUs training process.

\subsection{Neural Modules}
This part presents neural modules in OpenMatch, including \texttt{Embedder}, \texttt{Attention}, \texttt{Encoder}, and \texttt{Matcher}. All these flexible and extensible modules help to implement different Neu-IR models.

\textbf{Embedder.} The \texttt{Embedder} class mainly focuses on mapping tokens to dense representations. It feeds the token ids as inputs and returns a sequence of token embeddings for the following neural modules. The \texttt{Embedder} module can choose different \texttt{Tokenizer} and the token representations can be initialized with randomly initialized word embeddings, word2vec-based word embeddings, or subword embeddings from deep pretrained language models.

\textbf{Attention.} The \texttt{Attention} mechanism has been widely used in various deep learning tasks and is regarded as a crucial component in Neu-IR models. OpenMatch provides two kinds of attention modules in its library, including scaled dot-product attention and multi-head self-attention~\cite{vaswani2017attention}.

\textbf{Encoder.} The \texttt{Encoder} class provides various modules in deep learning to encode queries and documents, which consists of the feed-forward encoder, the CNN-based encoder, and multi-head self-attention based encoder. These \texttt{Encoder} modules further encode word representations to get their contextual representations. Users can set the dimension size, kernel size, and the number of the convolution kernel, the number of self-attention heads, and many other parameters in \texttt{Encoder} modules.

\textbf{Matcher.} The \texttt{Matcher} class is designed to measure the similarity between queries and documents by conducting interactions between them, which are used in previous Neu-IR models~\cite{macavaney2019cedr,dai2018convolutional,xiong2017end,hofstatter2020interpretable}. Following widely-used IR models~\cite{xiong2017end,dai2018convolutional}, OpenMatch implements a \texttt{KernelMatcher} module that employs RBF kernels~\cite{xiong2017end} to establish interactions between queries and documents and estimate their similarities with the term-term based soft matching.

\subsection{Few-Shot Training}
In reality, many ranking scenarios are few-shot and lack training data, especially for some specific domains, such as biomedical and legal search. Thus, broadening the benefits of deep neural networks to these few-shot scenarios is a standing challenge in information retrieval~\cite{roberts2020trec,yang2019critically}.
OpenMatch consists of two kinds of methods for few-shot IR tasks to alleviate data scarcity: using domain knowledge graphs and training Neu-IR models with large-scale selected weak supervision data. These methods incorporate human knowledge and massive relevance labels to better learn the semantics of queries and documents, which significantly alleviate the data dependency in few-shot ranking scenarios~\cite{zhang2020selective,sun2020meta}.

\textbf{Domain Knowledge Enhancement.} We follow the previous knowledge guided Neu-IR model EDRM~\cite{liu2018entity} and aim to incorporate the semantics from domain knowledge graphs to improve ranking performance in few-shot IR scenarios. EDRM comes up with enriched entity representations with the learned entity representations~\cite{Bordes2013TranslatingEF}, entity types and entity descriptions. OpenMatch leverages these technologies to enhance the semantic representations of queries and documents, and conduct better interactions between queries and documents for ranking.

\textbf{Data Augmentation.} Recent work~\cite{zhang2020selective,sun2020meta} provides several ways to augment training data, such as using anchor-document to approximate query-document relevance~\cite{zhang2020selective} and generating queries to synthesize relevance labels~\cite{sun2020meta}. To further boost model performance, they propose different strategies to select or reweight weak supervision data according to the performance of Neu-IR models on the target data. OpenMatch borrows these methods to guarantee the advance of Neu-IR models in few-shot scenarios.

\subsection{Training $\&$ Evaluation}

OpenMatch also provides the training, inference, and evaluation scripts for Neu-IR models. The detailed experimental instructions and results are all provided in OpenMatch, facilitating users and researchers to quickly start and easily reproduce the ranking results of different IR systems with OpenMatch.

\textbf{Training.} OpenMatch provides a common training framework for Neu-IR models. We define two widely-used learning-to-rank functions for users to optimize Neu-IR models, including point-wise training and pair-wise training.

\textbf{Evaluation.} We inherit the TREC evaluation tool, pytrec\_eval\footnote{https://github.com/cvangysel/pytrec\_eval}, to provide various standard metrics, such as NDCG and MAP. Besides, we also implement another widely-used evaluation metric MRR in OpenMatch. All above evaluation metrics are encapsulated in OpenMatch's \texttt{Metric} class. Users can evaluate their IR models with only one line of python command.

%% file: Sections/5_Model_Hub.tex
\section{Model Hub}
\label{sec:modelhub}
This section presents the model hub of OpenMatch. These models can be divided into retrieval models and reranking models according to their functions. As shown in Table~\ref{tab:model}, OpenMatch provides the implementation and detailed instructions of widely-used Neu-IR models for users. With well-defined model library, OpenMatch helps users quickly establish a Neu-IR model in their experiments, as shown in Figure~\ref{fig:example}.
\input{Tables/modelhub}

\textbf{Retrieval.} Given the user query and a large-scale document collection, retrieval models are used to select a subset from the document collection according to the given query.
OpenMatch consists of two kinds of retrieval models -- sparse models and dense models. The sparse models usually use discrete bag-of-word matching, such as BM25~\cite{robertson1995okapi} and SDM~\cite{metzler2007linear}. Lots of previous work incorporates different sparse retrieval features with learning-to-rank methods to implement a strong baseline~\cite{xiong2017end,dai2018convolutional,dai2019deeper}. In OpenMatch, we provide twenty sparse retriever methods to reproduce these baselines.

Recently, dense retrievers provide an opportunity to conduct the semantic matching supported by pretrained language models, such as DPR~\cite{karpukhin2020dense} and ANCE~\cite{xiong2020approximate}.
Moreover, to satisfy more complicated information needs with interactions between retrieval systems and users, OpenMatch 
also incorporates the conversational dense retriever (ConvDR)~\cite{yu2021convdr}, which inherits the document dense representation learned in ad hoc search and map conversational queries in the embedding space for retrieval. OpenMatch also provides APIs to calculate ranking scores with them.


\textbf{Reranking.} Reranking models in OpenMatch consist of Neu-IR models, pretrained IR models and learning-to-rank models, and aim to provide more accurate ranking results.
OpenMatch implements several typical and previously state-of-the-art Neu-IR models with its neural modules. For example, K-NRM~\cite{xiong2017end} can be built with \texttt{Embedder} and \texttt{KernelMatcher} modules. Conv-KNRM~\cite{dai2018convolutional} encodes queries and documents as n-grams representations with convolution layers and can be implemented with \texttt{Embedder}, \texttt{Conv1DEncoder}, and \texttt{KernelMatcher} modules. TK~\cite{hofstatter2020interpretable} uses multi-head attention layers as the encoder and can be implemented with \texttt{Embedder}, \texttt{TransformerEncoder}, and \texttt{KernelMatcher} modules. 
For pretrained Neu-IR models, we also inherit the pretrained language models from Huggingface's Transformers\footnote{\url{https://github.com/huggingface/transformers}} to implement reranking models. Several pretrained models can be used, such as BERT~\cite{devlin2018bert}, ELECTRA~\cite{clark2020electra}, and RoBERTa~\cite{liu2019roberta}. OpenMatch brings these latest pretrained language models from Huggingface's Transformers to Neu-IR models under the unified training framework.
Besides, OpenMatch adopts RankLib\footnote{\url{https://sourceforge.net/p/lemur/wiki/RankLib/}\label{url:ranklib}} to provide various learning-to-rank methods, such as RankNet~\cite{burges2005learning}, Coordinate Ascent~\cite{metzler2007linear}, LambdaMART~\cite{wu2010adapting}, and Random Forests~\cite{cutler2012random}. These learning-to-rank methods are usually utilized in previous work~\cite{dai2019deeper,zhang2020selective,sun2020meta} and aim to combine the ranking features from different IR models of both retrieval and reranking for further improvement.

\textbf{Ranking Results.} We also use our OpenMatch library to reimplement different Neu-IR models and conduct ranking results on several benchmarks, as shown in Table~\ref{tab:result}. Six datasets, ClueWeb09~\cite{callan2009clueweb09}, Robust04~\cite{kwok2004trec}, TREC COVID~\cite{wang2020cord}, MS MARCO~\cite{nguyen2016ms}, TREC CAsT-19~\cite{dalton2020cast} and TREC CAsT-20~\cite{cast2020overview}, are used in our experiments. With OpenMatch, we reproduce corresponding results of previous work on different ranking benchmarks.

\input{Tables/model_results}

%% file: Tables/modelhub.tex
\begin{table}[tbp]
    \centering
    \caption{OpenMatch Model Hub.}\label{tab:model}
    \begin{tabular}{l|l}
    \hline
    \textbf{Components} & \textbf{Models} \\
    \hline
    Sparse Retriever & \multicolumn{1}{m{5.5cm}}{Boolean AND/OR~\cite{salton1983extended}, LM~\cite{ponte1998language}, BM25~\cite{robertson1995okapi}, SDM~\cite{metzler2005markov}, TF-IDF~\cite{salton1988term}, Cosine Similarity, Coordinate match, Dirichlet LM ...} \\
    \hline
    Dense Retriever & \multicolumn{1}{m{5.5cm}}{DPR~\cite{karpukhin2020dense}, ANCE~\cite{xiong2020approximate}, ConvDR~\cite{yu2021convdr}} \\
    \hline
    Neural Ranker & \multicolumn{1}{m{5.5cm}}{K-NRM~\cite{xiong2017end}, Conv-KNRM~\cite{dai2018convolutional}, TK~\cite{hofstatter2020interpretable}, BERT~\cite{devlin2018bert}, RoBERTa~\cite{liu2019roberta}, ELECTRA~\cite{clark2020electra} ...} \\
    \hline
    LeToR & \multicolumn{1}{m{5.5cm}}{Coordinate Ascent~\cite{metzler2007linear}, RankNet~\cite{burges2005learning}, LambdaMART~\cite{wu2010adapting}, Random Forests~\cite{cutler2012random} ...} \\
    \hline
    
    
    \end{tabular}
\end{table}

%% file: Tables/model_results.tex
\begin{table}[tbp]
    \centering
    \caption{OpenMatch Experimental Results on Different Ranking Benchmarks. The ranking results from the corresponding work are reproduced. The ad hoc benchmark consists of two datasets, ClueWeb09 and Robust04.}\label{tab:result}
    \resizebox{0.99\linewidth}{!}{
    \begin{tabular}{l|c|c|c|c}
    \hline
    \multirow{2}{*}{\textbf{Models}} & \multicolumn{4}{c}{\textbf{Benchmarks}} \\
    \cline{2-5}
    & \textbf{Ad hoc} &\textbf{TREC COVID} & \textbf{MS MARCO}& \textbf{CAsT} \\
    \hline
    RankSVM~\cite{dai2019deeper} & \CheckmarkBold & \CheckmarkBold & \\
    Coor-Ascent~\cite{dai2019deeper} & \CheckmarkBold & \CheckmarkBold & \\
    K-NRM~\cite{xiong2017end} & \CheckmarkBold &  & \\
    ConvKNRM~\cite{dai2018convolutional,qiao2019understanding} & \CheckmarkBold &  & \CheckmarkBold \\
    EDRM~\cite{liu2018entity} & \CheckmarkBold &  & \\
    ReInfoSelect~\cite{zhang2020selective} & \CheckmarkBold & \CheckmarkBold & \\
    MetaAdaptRank~\cite{sun2020meta} & \CheckmarkBold & \CheckmarkBold & \\
    BM25 w. BERT~\cite{qiao2019understanding,dai2019deeper,yu2020few} & \CheckmarkBold & \CheckmarkBold & \CheckmarkBold & \CheckmarkBold \\
    ANCE w. BERT~\cite{xiong2020approximate,yu2021convdr} &  & & \CheckmarkBold & \CheckmarkBold \\
  
    \hline
    \end{tabular}}
\end{table}

%% file: Sections/2_Related_Work.tex
\section{Related Work}
Information Retrieval (IR) systems usually employ two-step pipelines to search related documents for user queries, which consist of retrieval and reranking. The IR community has a long history of building open-source toolkits for researchers. Anserini~\cite{yang2017anserini} is built on Lucene and aims to provide classical sparse retrieval methods that efficiently calculate query-document relevance according to the discrete term matching, such as BM25~\cite{robertson1995okapi} and SDM~\cite{metzler2005markov}. Pyserini~\cite{lin2021pyserini} also focuses on providing effective first-stage retrieval with spare retrieval and dense retrieval. MatchZoo~\cite{guo2019matchzoo} mainly focuses on conducting various neural text matching models beyond IR tasks~\cite{guo2019matchzoo}, which can be used in the reranking stage. Other toolkits, OpenNIR~\cite{macavaney2020opennir} and Capreolus\footnote{\url{https://github.com/capreolus-ir/capreolus}\label{url:cap}}, further help to build a whole IR pipeline with sparse retrieval and neural model based reranking. Some widely used Neu-IR models are incorporated, such as K-NRM~\cite{xiong2017end}, Conv-KNRM~\cite{dai2018convolutional}, CEDR~\cite{macavaney2019cedr}, and BERT~\cite{devlin2018bert}. However, these toolkits mainly focus on the architectures and variety of Neu-IR models rather than the problems that Neu-IR research usually faces, such as baseline implementation, model optimization and few-shot training. OpenMatch implements widely-used Neu-IR models, provides reproducible ranking results and advanced few-shot training methods, benefiting the Neu-IR research.

%% file: Sections/6_Conclusion.tex
\section{Conclusion}
With the development of deep neural networks, Neu-IR models have attracted lots of attention from the IR community. In practice, the performance of Neu-IR models determined by the model implementation, experimental details and training strategies. To help researchers implement standard Neu-IR models, OpenMatch provides different modules severed for data preprocessing, model building, Neu-IR model training, ranking feature ensemble and ranking performance evaluation. Using our OpenMatch, researchers can implement different neural/non-neural models for retrieval and reranking and establish tailored IR pipelines for experiments. Besides, OpenMatch guarantees researchers to reproduce ranking results of previous work by providing reproducible ranking results, experimental methodology and model implementation.


%% file: Sections/7_Acknowledge.tex
\section{Acknowledgements}
We thank Si Sun, Shi Yu, Yizhi Li and Aowei Lu as the contributors of OpenMatch. Part of this work is supported by the National Key Research and Development Program of China (No. 2020AAA0106501) and Beijing Academy of Artificial Intelligence (BAAI).